\documentclass[sigconf,nonacm]{acmart}

\usepackage{graphicx}
\usepackage{textcomp}
\usepackage{array}
\usepackage{caption}
\usepackage{multirow}
\usepackage{titlesec}
\usepackage{setspace}
\usepackage{algorithm}
\usepackage{algorithmic}
\usepackage{appendix}
\usepackage{listings}
\usepackage{color, colortbl}
\usepackage{amsmath}
\usepackage{enumitem}
\usepackage{tabularx}
\usepackage{tikz}
\usepackage{etoolbox}
\usepackage{stfloats}
\usepackage{color}
\usepackage{booktabs} 
\usetikzlibrary{matrix,positioning}
\usepackage{tabularray}
\usepackage{makecell}
\usepackage{lipsum}
\usepackage{tablefootnote}
\usepackage{subcaption}
\usepackage{tcolorbox}
\newtcolorbox{titleEnv}{
colframe=black!80,
colback=gray!10,
fonttitle=\bfseries,
coltitle=black,
left=3pt,
right=3pt,
top=3pt,
bottom=3pt,
boxrule=0.4mm,
arc=3mm
}

\definecolor{codegreen}{rgb}{0,0.6,0}
\definecolor{codegray}{rgb}{ .901,  .901,  .901}
\definecolor{codepurple}{rgb}{0.58,0,0.82}
\definecolor{backcolour}{rgb}{0.95,0.95,0.92}

\lstdefinestyle{mystyle}
{
    backgroundcolor=\color{backcolour},   
    commentstyle=\color{codegreen},
    keywordstyle=\color{magenta},
    numberstyle=\tiny\color{codegray},
    stringstyle=\color{codepurple},
    basicstyle=\ttfamily\scriptsize,
    breakatwhitespace=false,         
    breaklines=true,                 
    captionpos=b,                    
    keepspaces=true,                 
    numbers=left,                    
    numbersep=5pt,                  
    showspaces=false,                
    showstringspaces=false,
    showtabs=false,                  
    tabsize=2,
    xleftmargin=2em,
    xrightmargin=2em,
    frame=single,
    xleftmargin=0.03\textwidth, xrightmargin=-0.01\textwidth
}

\author{Shihan Dou}
\email{shdou21@m.fudan.edu.cn}
\authornote{Equal contribution}
\affiliation{%
  \institution{Fudan University}
  \city{Shanghai}
  \country{China}
}

\author{Junjie Shan}
\email{jshan@kth.se}
\authornotemark[1]
\authornote{Also with KTH Royal Institute of Technology}
\affiliation{%
  \institution{Westlake University}
  \city{Hangzhou}
  \country{China}
}

\author{Haoxiang Jia}
\email{haoxiangjia@hust.edu.cn}
\affiliation{%
  \institution{Huazhong University of Science and Technology}
  \city{Wuhan}
  \country{China}
}

\author{Wenhao Deng}
\email{wenhao.deng@foxmail.com}
\affiliation{%
  \institution{Westlake University}
  \city{Hangzhou}
  \country{China}
}

\author{Zhiheng Xi}
\email{zhxi22@m.fudan.edu.cn}
\affiliation{%
  \institution{Fudan University}
  \city{Shanghai}
  \country{China}
}

\author{Wei He}
\email{whe23@m.fudan.edu.cn}
\affiliation{%
  \institution{Fudan University}
  \city{Shanghai}
  \country{China}
}

\author{Yueming Wu}
\email{wuyueming21@gmail.com}
\authornote{Yueming Wu is the corresponding author}
\affiliation{%
  \institution{Nanyang Technological University}
  \country{Singapore}
}

\author{Tao Gui}
\email{tgui@fudan.edu.cn}
\affiliation{%
  \institution{Fudan University}
  \city{Shanghai}
  \country{China}
}

\author{Yang Liu}
\email{yangliu@ntu.edu.sg}
\affiliation{%
  \institution{Nanyang Technological University}
  \country{Singapore}
}

\author{Xuanjing Huang}
\email{xjhuang@fudan.edu.cn}
\affiliation{%
  \institution{Fudan University}
  \city{Shanghai}
  \country{China}
}

\usepackage{url}

\newcommand{\ie}{\textit{i.e.,} }
\newcommand{\eg}{\textit{e.g.,} }

\begin{document}

\title{Towards Understanding the Capability of Large Language Models on Code Clone Detection: A Survey}

\begin{abstract}

Code cloning, the duplication of code fragments, is common in software development. 
While some reuse aids productivity, excessive cloning hurts maintainability and introduces bugs. 
Hence, automatic code clone detection is vital. 
Meanwhile, large language models (LLMs) possess diverse code-related knowledge, making them versatile for various software engineering challenges. 
However, LLMs' performance in code clone detection is unclear and needs more study for accurate assessment.
In this paper, we provide the first comprehensive evaluation of LLMs for clone detection, covering different clone types, languages, and prompts. 
We find advanced LLMs excel in detecting complex semantic clones, surpassing existing methods. 
Adding intermediate reasoning steps via chain-of-thought prompts noticeably enhances performance. Additionally, representing code as vector embeddings, especially with text encoders, effectively aids clone detection.
Lastly, the ability of LLMs to detect code clones differs among various programming languages.
Our study suggests that LLMs have potential for clone detection due to their language capabilities, offering insights for developing robust LLM-based methods to enhance software engineering.

\end{abstract}



\maketitle
\settopmatter{printfolios=true}

\section{Introduction}

Code cloning, the replication of code fragments, is a common phenomenon in software development. While some code reuse aids productivity, excessive cloning negatively impacts maintainability and propagates bugs \cite{hummel2010index, kim2017vuddy}. 
Thus, automatic clone detection is an important research area. 
To better comprehend clone detection, researchers have undertaken a methodical classification of code clones into distinct categories. 
A widely accepted taxonomy segregates code clones into four types: Type-1 (identical similarity), Type-2 (lexical similarity), Type-3 (syntactical similarity), and Type-4 (semantic similarity) \cite{bellon2007type1_4, roy2007type1_4}. 
The first three types can generally be encapsulated under the umbrella of syntactic similarities, while the fourth type epitomizes semantic similarities. 
Given that Type-4 clones may comprise clones that display a wide range of syntactic dissimilarities, they present the most formidable challenge for most clone detection methodologies.
There exists extensive literature focusing on code syntactic similarities \cite{sajnani2016sourcerercc, roy2008nicad, nakagawa2021nil}. 
However, in recent years, attention has gradually shifted toward the study of code semantic similarities. 
This shift has been facilitated by advancements in the field of deep neural networks. 
As a result, a plethora of deep learning-based methodologies have been proposed, all designed to discern semantic similarities through a process of data-driven learning \cite{lei2022deep}. 
These methodologies largely adopt a two-pronged approach: firstly, neural networks are leveraged to generate a vector representation for each code fragment, which is then followed by calculating the similarities between the vector representations of two code fragments to detect clones \cite{wu2020scdetector}.


As a matter of fact, the development of pre-trained language models (PLMs) has revolutionized the area of deep learning.
These models, such as BERT \cite{kenton2019bert} and GPT-1 \cite{radford2018improving}, were pre-trained with specially designed pre-training tasks on large-scale unlabeled text corpora to learn generalized knowledge.
After that, many works such as CodeBERT \cite{feng2020codebert} and CodeT5+ \cite{wang2023codet5+} introduce pre-training to further boost code-related tasks in software engineering.
Although these works have a great performance, they still need to be fine-tuned to adapt to different downstream tasks \cite{sun2019fine, liu2023pre}.
Recently, researchers have found that scaling PLMs (\eg scaling model size or data size) often leads to an improved model capacity on downstream tasks \cite{kaplan2020scaling}.
Although scaling is mainly conducted in model size with similar architectures and pre-training tasks, these large-sized PLMs (\eg GPT-3 \cite{brown2020language}, MPT \cite{MosaicML2023Introducing}, LLaMA \cite{touvron2023llama}) display different behaviors from smaller PLMs (\eg 330M-parameter BERT and 1.5B-parameter GPT-2 \cite{radford2019language}) and show surprising abilities in solving a series of complex tasks with only human instructions rather than fine-tuning to adapt the downstream tasks \cite{wei2022emergent, brown2020language}. 
Furthermore, since the pre-trained corpus of these large language models (LLMs) contains a huge amount of code tasks, they are also enabled to solve a variety of challenges related to code in software engineering.
For example, Feng \emph{et al.}~\cite{feng2023prompting} proposed an automatic technique for accomplishing the bug replay from bug reports through prompt engineering.
Deng \emph{et al.}~\cite{deng2023large} proposed a testing tool, using generative and infilling LLMs to generate and mutate various programs for testing the deep learning library.
However, there is a lack of understanding of how well these LLMs perform in code clone detection.

In our paper, we delve into the potential of leveraging LLMs for detecting code clones.
Our hypothesis pivots on the innate ability of LLMs to interpret complex language inputs and generate meaningful outputs. 
We posit these skills could be harnessed to identify and classify code clones, thus providing a novel approach to a traditional code clone detection problem.
Specifically, we conduct a comprehensive study to assess the clone detection performance of LLMs like Llama \cite{touvron2023llama}, Alpaca \cite{taori2023alpaca}, Vicuna \cite{zheng2023judging}, StarChat-$\beta$ \cite{Tunstall2023starchat-alpha}, Falcon \cite{almazrouei2023falcon}, MPT \cite{MosaicML2023Introducing}, Llama2 \cite{touvron2023llama2}, Llama2-Chat \cite{touvron2023llama2}, GPT-3.5 \cite{ouyang2022training}, and GPT-4 \cite{openai2023gpt4}. 
Our study focuses on the following research questions:
\begin{itemize}
\item \emph{RQ1: Can LLMs detect code clones with a simple prompt?}
\item \emph{RQ2: How do LLMs perform by using one-step chain-of-thought prompts?}
\item \emph{RQ3: Can LLMs perform better by using multi-step chain-of-thought prompts?}
\item \emph{RQ4: How do LLMs perform using code embedding?}
\item \emph{RQ5: How does the performance of LLMs in code clone detection vary across different programming languages?}
\end{itemize}

Regarding \textbf{RQ1}, our findings indicate that when utilizing only a simple prompt, clone detection based on open-source LLMs performs better in detecting Type-3 and Type-4 clone pairs compared to existing tools. 
However, it performs slightly worse in detecting Type-1 and Type-2 clone pairs.
GPT-3.5-Turbo and GPT-4 have the highest recall and accuracy in almost all clone types. 
Regarding \textbf{RQ2}, our observations reveal that employing one-step chain-of-thought reasoning significantly enhances the performance of GPT-3.5-Turbo and GPT-4. 
This improvement is attributed to the intermediate reasoning, which allows the larger models to consider the code from multiple perspectives, resulting in more accurate clone detection.
Surprisingly, when incorporating all the intermediate reasoning together, GPT-3.5-Turbo's effectiveness decreases, and it even performs worse than when using a simple prompt. 
In contrast, GPT-4's detection remains unaffected by this integration.
Regarding \textbf{RQ3}, when multiple reasonings are generated simultaneously, we observe that the reasoning from different angles can interfere with each other, leading to a decrease in the detection results. 
Moreover, we also conduct simulations of deep learning-based clone detection by independently generating code explanations for each code pair. 
This approach yields positive results and can achieve more accurate and reliable clone detection outcomes.
Regarding \textbf{RQ4}, when it comes to code embedding, Text-embedding-ada-002 is more effective than specialized CodeBERT models in identifying cloned code, exhibiting superior overall performance.
Regarding \textbf{RQ5}, we discover that the effectiveness of LLMs in detecting code clones varies across different programming languages, with Python generally producing better results, probably because it is naturally simple and frequently used in training data.


In summary, our paper makes the following contributions:
\begin{itemize}
    \item We perform the first empirical study to assess the capability of existing LLMs in detecting code clones from five different perspectives (\ie simple prompts, one-step chain-of-thought prompts, multi-step chain-of-thought prompts, code embedding, and multiple programming languages).
    \item We open source all the data and code involved in our study and offer valuable insights into the capabilities and limitations of LLMs for code clone detection.
    The results obtained will serve as essential guidance for future research aimed at improving LLM-based clone detection and other aspects of software engineering.
\end{itemize}


\par \noindent \textbf{Paper Organization.} The remainder of the paper is organized as follows.
Section 2 explains the background.
Section 3 introduces our experimental setup. 
Section 4 reports the experimental results. 
Section 5 discusses future work. 
Section 6 concludes the present paper.

\section{Background and Related Work}
In this section, we briefly introduce clone code detection, Large Language Models (LLMs), and chain-of-thought reasoning.

\subsection{Code Clone Detection}
Code clone detection aims to dig out code snippets with similar functionalities, which has attracted wide attention in software engineering~\cite{10.5555/832303.836911, Kim2005AnES, roy2008nicad}.
Commonly, code clone types are classified into four categories based on syntactic or semantic differences~\cite{bellon2007type1_4}. 
\textbf{Type-1 (identical similarity)} refers to identical code fragments, differing only in white-space, layout, and comments.
\textbf{Type-2 (lexical similarity)} entails identical code fragments with variations in identifier names and lexical values, in addition to the differences present in Type-1 clones.
\textbf{Type-3 (syntactic similarity)} consists of syntactically similar code snippets that vary at the statement level. In addition to the differences found in Type-1 and Type-2 clones, these fragments have statements added, modified, and/or removed with respect to each other.
\textbf{Type-4 (semantic similarity)} refers to syntactically dissimilar code fragments that implement the same functionality. 

Many approaches have been proposed to detect code clones, they can be broadly categorized into various types, including text-based \cite{johnson1994substring, ducasse1999language, roy2008nicad, ragkh2017using, kim2018software, jadon2016code, yu2017detecting}, token-based \cite{gode2009incremental, kamiya2002ccfinder, sajnani2016sourcerercc, li2017cclearner, wang2018ccaligner, golubev2021multi, hung2020cppcd}, tree-based \cite{jiang2007deckard, wei2017cdlh, zhang2019astnn, liang2021astpath, jo2021twopass, pati2017comparison, chodarev2015haskell, hu2022treecen, wu2022amain}, and graph-based \cite{krinke2001duplix, komondoor2001pdgdup, wang2017ccsharp, zhao2018deepsim, wu2020scdetector, zou2020ccgraph} tools. 
Moreover, since the automatic feature extraction of deep learning, it is also being increasingly adopted for cloned code detection tasks by processing different code representations \cite{wei2017cdlh, zhang2019astnn, hu2022treecen, wu2022amain, zhao2018deepsim, wu2020scdetector}.
However, since the rapid development of large language models, there has been no work to detect cloned code by using large language models, and there has been no more thorough exploration of the performance of large language models for detecting code clones.

\subsection{Large Language Models}
The recent advancements in Large Language Models (LLMs) have sparked a revolution in Natural Language Processing (NLP).
In general, a large language model is a Transformer-based model containing hundreds of billions (or more) of parameters, such as LLaMA~\cite{touvron2023llama}, Vicuna~\cite{zheng2023judging}, Falcon~\cite{almazrouei2023falcon}, StarChat-$\beta$ \cite{Tunstall2023starchat-alpha} and GPT4~\cite{openai2023gpt4}. 
These models, trained on a massive corpus of text, have the ability to learn a vast array of knowledge from the text, thereby tackling a multitude of complex tasks in NLP and understanding human queries to engage in unbounded dialogues.


In the context of earlier language sequence tasks, including both natural and programming languages, satisfactory performance has been achieved through task-specific fine-tuning \cite{sun2019fine}. 
Fine-tuning is the process of updating model weights by learning the relationship between input and output from a specific downstream task dataset \cite{liu2023pre}. 
However, given the comprehensive knowledge encapsulated within LLMs, a novel method, known as In-context Learning \cite{dai2022can}, can be utilized to apply LLMs to downstream tasks. 
In contrast to fine-tuning, which typically necessitates large downstream datasets for model tuning, in-context learning enables LLMs to understand tasks through instructions and examples, leveraging their inherent capabilities \cite{wei2022emergent, dong2022survey}.
In this study, we developed a variety of instructions to guide LLMs to understand the task of clone code detection from multiple perspectives, thereby facilitating a comprehensive evaluation of the LLMs' performance on code clone detection.

\subsection{Chain-of-Thought Reasoning}

Traditional small language models typically struggle to solve complex tasks or answer difficult questions that involve multiple reasoning steps, such as mathematical word problems. 
By contrast, LLMs, employing the chain-of-thought (CoT) prompting strategy~\cite{wei2022chain}, can address these tasks or dissect complex problems by using an intermediate reasoning process to derive the final answer.
CoT prompting, distinct from the traditional direct-answer prompt, enables the model to formulate a thought process for the question before providing an answer. 
Alternatively, it can manually decompose a complex question into multiple intermediate steps for the model to resolve. 
This approach, similar to human cognitive processes, can enhance the performance of large models when faced with complex problems. 
A number of studies \cite{li2022advance, fu2022complexity, zhang2023multimodal} have demonstrated that CoT prompting can yield significant performance gains in complex reasoning benchmarks.



Given the proven efficacy of CoT prompting in increasing the accuracy of complex problem resolution by introducing intermediate reasoning steps, this paper aims to investigate the performance of CoT in the task of cloned code detection, both from one-step and multi-step perspectives. 
In one-step prompt engineering, the model is tasked with detecting code clones from various perspectives (\ie clone type, similarity, and analogous lines of code pair). 
In multi-step prompt engineering, the model initially analyzes each function from multiple perspectives, subsequently integrating all the intermediate reasonings. 
This approach enables the model to detect code clones with prior knowledge, rather than merely following human instructions to provide a binary "yes" or "no" response.

\section{Experimental Setup}

\begin{table*}[]
\centering
\caption{Prompt Design for Code Clone Detection Research Questions 1\textasciitilde5}
\label{prompt-1}
\begin{spacing}{1}
\small
\setlength{\tabcolsep}{0.65mm}{
\centering
\begin{tabular}{c|c|m{14.3cm}}
\toprule
\toprule
\textbf{RQ}    & \textbf{Instruction Type} & \textbf{Instance} \\
\midrule
\textbf{1}     & Simple Prompt & Please analyze the following two code snippets and determine if they are code clones. Respond with `yes' if the code snippets are clones or `no' if not. \\
\midrule
\multirow{5}[2]{*}{\textbf{2}} & Clone Type     &  Please analyze the following two code snippets and determine if they are code clones. Respond with `yes' if the code snippets are clones or `no' if not. If the answer is yes, please report the specific clone type (\ie Type-1, Type-2, Type-3, or Type-4). \\ \cmidrule{2-3}
& Similarity  & Please assess the similarity of the following two code snippets and provide a similarity score between 0 and 10. A higher score indicates that the two codes are more similar. Output the similarity score. \\ \cmidrule{2-3}
& Reasoning     & Please provide a detailed reasoning process for detecting code clones in the following two code snippets. Based on your analysis, respond with `yes' if the code snippets are clones or `no' if they are not. \\ \cmidrule{2-3}
& Similar Line     & Please analyze the following two code snippets for code clone detection. You should first report which lines of code are more similar. Then based on the report, please answer whether these two codes are a clone pair. The response should be `yes' or `no'. \\ \cmidrule{2-3}
& Integrated     & Please analyze the following two code snippets to assess their similarity and determine if they are code clones. Provide a similarity score between 0 and 10, where a higher score indicates more similarity. Additionally, identify the type of code clone they represent and present a detailed reasoning process for detecting code clones. \\
\midrule
\multirow{2}[2]{*}{\textbf{3}} 
& Separate Explanations     & \colorbox{codegray}{\textbf{Step1:}} The same as RQ2's prompt without the final code clone detection judgment. \newline \colorbox{codegray}{\textbf{Step2:}} Please analyze the following two code snippets and determine if they are code clones. The Clone Type/Similarity/Reasoning/Difference/Integrated information of the first and the second code is \textbf{\{Step1 Output\}}. Please respond with `yes' if the code snippets are clones or `no' if they are not.
\\ 
\cmidrule{2-3}
& Separate Codes     & \colorbox{codegray}{\textbf{Step1 \& 2:}} Please analyze the following code snippet and explain the function of the snippet. \newline \colorbox{codegray}{\textbf{Step3:}} Please analyze the following two code snippets and determine if they are code clones. The function of the first code is \textbf{\{Step1 Output\}} and the second is \textbf{\{Step2 Output\}}. Please answer `yes' if the code snippets are clones or `no' if they are not.\\  

\midrule
\textbf{5}     & Simple Prompt & Same as RQ1. \\

\bottomrule
\bottomrule

\end{tabular} }%
\end{spacing}
\end{table*}

\subsection{Research Questions}
Our empirical study delved into five research questions to improve the understanding of code clone detection using LLMs. 
\begin{itemize}[leftmargin=8pt]
    \item \textbf{RQ1: Can LLMs detect code clones with a simple prompt?}
    We aim to explore the performance of LLMs in code clone detection tasks under these conditions. 
    Specifically, we design a prompt to ask LLMs to answer the code clone detection judgment directly, expecting them to output a simple "Yes" or "No". 
    This facilitates data analysis across different clone types.

    \item \textbf{RQ2: How do LLMs perform by using one-step chain-of-thought prompts?} 
    Given the inherent nature of language models as posterior probability estimators, we intend to improve LLM performance by altering instructions for various perspectives. Specifically, we design prompts to direct the model to conduct code analysis prior to the code clone detection judgment. The code analysis encompasses five techniques: clone type discrimination, similarity calculation, reasoning explanation, similar line discrimination, and integrated analysis.
    
    \item \textbf{RQ3: Can LLMs perform better by using multi-step chain-of-thought prompts?} 
    While we have directed the model to analyze clone code from one or several perspectives in RQ2, language models may be influenced by other factors during code analysis, including the counterpart code in a code pair or different analysis angles. 
    So, we design prompts based on chain-of-thought reasoning and categorize them into two types: \textit{separate explanations} and \textit{separate code}. 
    The former prompts the LLMs to output the same code analysis information as in RQ2, and then, we request the LLMs, based on this output, to independently execute the code clone detection. 
    The latter prompts the LLMs to independently explain each code snippet's function. 
    Then, based on these outputs, we ask the LLMs to conduct code clone detection independently. 
    The ultimate goal is to enable the model to independently analyze each code in the pair or from various perspectives, aggregate the analysis results, and apply these findings to perform the final clone code detection more accurately.
    
    \item \textbf{RQ4: How do LLMs perform using code embedding?} 
    This question focuses on whether LLMs can provide superior results compared to traditional pre-trained language models (PLMs) through code compression. 
    We compare the performance of LLMs with specific models such as CodeBERT-base, CodeBERT-mlm, and text-embedding-ada-002.
    This comparison leverages the embedding API provided by OpenAI \cite{openai}.
    Since this research question primarily compares the performance of existing embedding models, we do not design specific prompts for it.
    
    \item \textbf{RQ5: How does the performance of LLMs in code clone detection vary across different programming languages?} 
    We aim to discern whether LLMs exhibit different performances in code clone detection across various programming languages. 
    For a fair comparison, we apply the prompts from RQ1 without specifying the language of the target code snippets.
    This allows the assessment of LLMs' versatility in handling diverse programming languages.

\end{itemize}

\subsection{Instructions}
We design different prompts to inspire the ability of large language models. Examples of the prompts are displayed in Table~\ref{prompt-1}.

\subsection{Dataset Collection}





Our evaluations were conducted using the BigCloneBench dataset \cite{big}, a comprehensive collection of over 8 million labeled clone pairs derived from 25,000 systems. 
Each clone pair in BigCloneBench corresponds to function-level code and is manually assigned an appropriate clone type.
Clone types are divided into Type-1 and Type-2, with additional sub-categories for Type-3 and Type-4 clones based on their syntactical similarity scores. 
These include i) \emph{Very Strongly Type-3} (VST3) clones, with similarity scores in the range of [0.9, 1.0); ii) \emph{Strongly Type-3} (ST3) clones, with similarity scores between [0.7, 0.9); iii) \emph{Moderately Type-3} (MT3) clones, with similarity scores between [0.5, 0.7); and iv) \emph{Weakly Type-3/Type-4} (WT3/T4) clones, with similarity scores between [0.0, 0.5).

In addition to Java, our study also included C/C++ and Python programming languages.
For these languages, we derived datasets from CodeNet \cite{puri2021codenet}, incorporating C++ and Python benchmarks. 
As the clone types in the C++ and Python benchmarks were not pre-classified in CodeNet, we conducted a classification following the standards set by BigCloneBench. 
This involved the use of respective lexical analyzers for Python and C++ code tokenization, after which we calculated Jaccard indices to measure syntactical similarity scores. 
Based on these scores, we categorized the code clones for each language, thus constructing a comprehensive and diverse code clone detection dataset for different programming languages.

To ensure a robust and comprehensive evaluation across all considered programming languages, we meticulously sampled our datasets. 
From the BigCloneBench dataset, we sampled 500 pairs of code for each clone type and included 3000 non-clone samples. 
For the C++ and Python languages, we sampled 100 pairs of code for each clone type and supplemented these with 600 non-clone samples. 
This diverse sampling was conducted while strictly adhering to the constraints of our available GPU computing resources.

\subsection{Language Models}

We evaluated 12 language models, including a variety of locally deployable open-source models, API-based LLMs, an LLM-generated code embedding model, and pre-trained language models for code embedding.

\subsubsection{Open-source Large Language Models} 
\ 
\newline
Eight of the models we evaluated are open-source LLMs, capable of local deployment. These include LLaMA \cite{touvron2023llama}, Alpaca \cite{taori2023stanford}, Vicuna \cite{zheng2023judging}, Falcon\cite{almazrouei2023falcon}, MPT \cite{MosaicML2023Introducing}, LLaMA2 \cite{touvron2023llama2}, LLaMA2-Chat \cite{touvron2023llama2}, and StarChat-$\beta$ \cite{Tunstall2023starchat-alpha}.
Each of these models has been trained on large corpora comprising both text and code, with parameters in the range of billions. 
These models are used to leverage their large-scale learning capability for code clone detection.

\textbf{LLaMA \cite{touvron2023llama} and LLaMA2 \cite{touvron2023llama2}:} 
LLaMA and LLaMA2 are large language models that have been trained on a corpus incorporating trillions of tokens. This corpus includes both text and code. Both models exhibit remarkable performance across various benchmarks, underlining their reliability. For our experiments, we deployed the 7-billion-parameter version of LLaMA, referred to as LLaMA-7B. 
LLaMA2, on the other hand, has been subjected to a more rigorous cleaning process during training and has consistently shown fantastic results on open benchmarks \cite{touvron2023llama2}.
Both LLaMA models represent the robustness and efficacy of large-scale language models in dealing with diverse and complex tasks \cite{10.1145/3605943, zhao2023survey}.

\textbf{Alpaca \cite{taori2023stanford}:} Alpaca is a unique language model that has been fine-tuned on LLaMA-7B. The fine-tuning process utilized approximately 52k instruction data. Alpaca's distinctive strength lies in its ability to follow instructions superior to its base model, LLaMA, thereby amplifying its performance on intricate tasks \cite{wang2023visionllm}.

\textbf{Vicuna \cite{zheng2023judging}:} Vicuna is another model that is built upon LLaMA-7B. 
Its fine-tuning process incorporates 70k user-shared multi-round conversations along with long-sequence samples. 
Like Alpaca, Vicuna exhibits an enhanced ability to comply with human instructions as compared to the original LLaMA model, providing it with a competitive edge to handle complex tasks \cite{zhao2023survey}.

\textbf{LLaMA2-Chat \cite{touvron2023llama2}:} LLaMA2-Chat is an open-source dialogue large language model, fine-tuned and aligned by Reinforcement Learning with Human Feedback (RLHF) \cite{ouyang2022training} based on LLaMA2, and achieves a great performance among open-source models on the human instruction benchmark. 
Except that the base model is different from Alpaca and Vicuna, LLaMA2-Chat also aligns with human feedback on the helpful and harmless data, which makes the model better able to understand human instructions, improving its usefulness and mitigating harmfulness \cite{zheng2023secrets, bai2022training}.

\textbf{Falcon-Instruct \cite{almazrouei2023falcon}:} Falcon-Instruct constitutes our list of evaluated open-source large language models. 
Falcon's uniqueness stems from its pre-training on a distinct corpus, complemented by a stringent cleaning process. 
Meanwhile, Falcon has also been trained on longer sequences, which can be expected to better address long content tasks such as code clone detection.
Falcon-Instruct fine-tuned based on Falcon has consistently demonstrated remarkable performance on a variety of open benchmarks \cite{li2023cmmlu}.

\textbf{MPT-Instruct \cite{MosaicML2023Introducing}:} MPT-Instruct is another open-source large language model we evaluated. 
MPT such as Falcon, it has been trained on a unique corpus and has undergone a rigorous cleaning process. 
As an open-source LLm, MPT-Instruct instruct-tuned based on MPT also has demonstrated strong performance across several open benchmarks, further validating its effectiveness \cite{touvron2023llama2}.

\textbf{StarChat-\boldmath{$\beta$} \cite{Tunstall2023starchat-alpha}:} StarChat-$\beta$ is a large language model that is instruction-tuned on an "uncensored" variant of the openassistant-guanaco dataset \footnote{\href{https://huggingface.co/datasets/timdettmers/openassistant-guanaco}{https://huggingface.co/datasets/timdettmers/openassistant-guanaco}} to act as a helpful coding assistant.
The base model of StarChat-beta is StarCoderPlus \cite{li2023starcoder}, which is a 15.5B parameter Language Model trained on English and more than 80 programming languages. 
Therefore StarChat-$beta$ is well capable of understanding human instructions while performing a variety of coding tasks.

\subsubsection{OpenAI Large Language Models}
\ 
\newline
We also assessed the performance of two OpenAI LLMs, GPT-3.5-turbo \cite{ouyang2022training} and GPT-4 \cite{openai2023gpt4}, that are accessible via their API. 
These advanced iterations of the GPT series language models provided by OpenAI have shown superior performance on a wide array of natural language processing and programming language tasks \cite{hendrycks2020measuring, liu2023jailbreaking}.

\subsubsection{Pre-trained Language Models for Code Embedding} 
\ 
\newline
Embedding is a machine learning technique that effectively converts high-dimensional and complex data, such as text and images, into simpler, lower-dimensional representations.  
Such representations can either be employed directly as feature representations or further refined using training data from subsequent supervised tasks.
We evaluated two models specifically designed for code embeddings: CodeBERT-Base \cite{feng2020codebert} and CodeBERT-MLM \cite{feng2020codebert}. 
CodeBERT-Base is trained on a mix of natural language and code corpora, whereas CodeBERT-MLM leverages a masked language modeling objective, enhancing its suitability for tasks that require understanding and analyzing code \cite{zhang2022survey}.
On the other hand, we also evaluated Text-embedding-ada-002 \cite{embeddingadav2}, which is an LLM-generated text embedding model that generates embeddings for natural language and code, making it particularly suitable for tasks such as code clone detection.

\subsubsection{Implementation}
\ 
\newline
When addressing code tasks by using a language model, most scenarios need to ensure accuracy rather than diversity of model responses, so we need to set the hyperparameters differently from the natural language task \cite{openai2023gpt4}. 
In all of our experiments, we set the temperature \cite{ackley1985learning, ficler-goldberg-2017-controlling}, Top-p (\ie Nucleus Sampling \cite{holtzman2019curious}), and Top-p \cite{fan2018hierarchical} of the inference phase to 0.2, 0.1 and 10, respectively.

\subsection{Non-LLMs-Based Detection Techniques}
We also select eight state-of-the-art code clone detection tools as baseline methods. 
SourcererCC \cite{sajnani2016sourcerercc} is a token-based clone detector that uses an inverted index data structure to swiftly query proportional clones of a given code block, detecting Type-1, Type-2, and Type-3 clones with high precision and recall. 
CCFinder \cite{kamiya2002ccfinder}, developed by Kamiya et al., is a four-phase detection tool based on a suffix tree-matching algorithm, capable of identifying clone pairs and classes of clones. 
NiCad \cite{roy2008nicad}, primarily employed for Android malware detection, is a text-based detector that utilizes Java source code to detect Type-1, Type-2, and Type-3 clones. 
Deckard \cite{jiang2007deckard}, a tree-based detector, converts source code into an abstract syntax tree and computes clone similarity through the comparison of characteristic vectors. 
CCAligner \cite{wang2018ccaligner}, another token-based detector, works with C and Java files to detect Type-1, Type-2, and Type-3 clones. 
Oreo \cite{saini2018oreo} presents a novel approach that combines machine learning, information retrieval, and software metrics to detect Type-1 to Type-3 clones and those in the Twilight Zone. 
LVMapper \cite{wu2020lvmapper} introduces an innovative detection approach for large-variance clones borrowed and adapted from sequencing alignment in bioinformatics, demonstrating an impressive recall for general Type-1, Type-2, and Type-3 clones. 
Lastly, NIL \cite{nakagawa2021nil} proposes a scalable token-based detection technique capable of identifying clone candidates efficiently using an N-gram representation of token sequences and an inverted index and is particularly proficient in detecting large-variance clones and ensuring scalability.

\subsection{Evaluation Metrics}
We use the following widely used metrics to measure the detection performance.
Precision is defined as $P = TP/(TP+FP)$.
Recall is defined as $R = TP/(TP+FN)$.
F1 is defined as $F1 = 2*P*R/(P+R)$.
Among them, \emph{true positive} (TP) represents the number of samples correctly classified as clone pairs, \emph{false positive} (FP) represents the number of samples incorrectly classified as clone pairs, and \emph{false negative} (FN) represents the number of samples incorrectly classified as non-clone pairs.

\subsection{Hardwares}
The experiments were conducted on a server equipped with dual AMD EPYC 7742 64-Core Processors, 128 CPUs, 1TB of memory, and eight NVIDIA A800-SXM4-80GB GPUs.
\section{Experimental Result}


\subsection{RQ1: Performance of A Simple Prompt}

\begin{table}[]
\centering
\caption{Comparison of SOTA Code Clone Detection Methods and LLMs-based Code Clone Detection Methods}
\label{tab:rq1-1}
\begin{spacing}{1.}
\small
\setlength{\tabcolsep}{1.mm}{

\begin{tabular}{c|ccccccc}
\toprule
\toprule
\multirow{2}{*}{\textbf{Methods}} & \multicolumn{6}{c}{\textbf{Recall}}   & \multirow{2}{*}{\textbf{Precision}} \\ 
\cmidrule{2-7} & \textbf{T1}  & \textbf{T2}   & \textbf{VST3}  & \textbf{ST3}  & \textbf{MT3} & \textbf{T4} &                                   \\ 

\midrule
\rowcolor[rgb]{ .901,  .901,  .901}\multicolumn{8}{l}{\textbf{Non-LLMs-Based Detection}} \\
\midrule
SourcererCC               & 1 & 0.97 & 0.93 & 0.60 & 0.05 & 0    & 0.98                             \\ 
CCFinder                 & 1 & 0.93 & 0.62 & 0.15 & 0.01 & 0   & 0.72                              \\
NiCad               & 1 & 0.99 & 0.98 & 0.93 & 0.008 & 0    & 0.99                              \\ 
Deckard                 & 0.6 & 0.58 & 0.62 & 0.31 & 0.12 & 0.01   & 0.35                              \\
CCAligner               & 1 & 0.99 & 0.97 & 0.70 & 0.1 & -    & 0.80                              \\ 
Oreo                 & 1 & 0.99 & 1 & 0.89 & 0.30 & 0.007   & 0.90                              \\
LVMapper               & 0.99 & 0.99 & 0.98 & 0.81 & 0.19 & -    & 0.58                              \\ 
NIL                 & 0.99 & 0.96 & 0.93 & 0.67 & 0.10 & -   & 0.94                             \\
\midrule
\rowcolor[rgb]{ .901,  .901,  .901}\multicolumn{8}{l}{\textbf{LLMs-Based Detection}} \\
\midrule
LLaMA-7B\footnotemark[2]               & - & - & - & - & - & -    & -                              \\
LLaMA2-7B\footnotemark[2]              & - & - & - & - & - & -    & -                                \\
Aplaca-7B                 & 0.76 & 0.93 & 0.65 & 0.87 & 0.89 & 0.71   & 0.55                             \\
Vicuna-7B               & 0.42 & 0.3 & 0.72 & 0.74 & 0.90 & 0.60    & 0.45                              \\
LLaMA2-Chat-7B                 & 1 & 1 & 0.998 & 1 & 1 & 0.990   & 0.51                              \\
Falcon-Instruct-7B               & 0.998 & 1 & 1 & 1 & 1 & 0.991   & 0.48                              \\
MPT-Instruct-7B               & 0.47 & 0.08 & 0.23 & 0.33 & 0.28 & 0.15    & 0.74                              \\  
StarChat-$\beta$-16B               & 0.93 & 0.49 & 0.42 & 0.43 & 0.26 & 0.37   & 0.62                              \\
GPT-3.5-Turbo               & 1 & 0.57 & 0.85 & 0.78 & 0.59 & 0.09    & 0.95                              \\ 
GPT-4                 & 1 & 0.98 & 0.99 & 0.94 & 0.77 & 0.15   & 0.96                              \\
\bottomrule
\bottomrule
\end{tabular} }%
\end{spacing}
\end{table}
\footnotetext[2]{ indicates base models that are not fine-tuned on instruction datasets, $-$ indicates the model did not return meaningful results.}

In this research question, we want to determine whether LLMs can conduct code clone detection with a simple prompt. We evaluate ten LLMs and eight non-LLMs-based code clone detection techniques on various datasets, including six clone types. From Table~\ref{tab:rq1-1}, we can observe that for Type-1 and Type-2 clones, non-LLMs-based detection tools have higher recall than LLMs-based detection tools, while LLMs-based detection tools perform better for Type-3 and Type-4 clones. 
Specifically, SourcererCC, NiCad, CCAligner, Oreo, and NIL show strong recall in T1, T2, and VST3 clones, with NiCad and Oreo also showing high recall for ST3 clones. 
However, for MT3 and T4 clones, these tools have significantly lower recall, indicating they may struggle with more complex or subtle forms of code duplication.
CCFinder and Deckard show lower recall across the board compared to the previous group, especially with ST3, MT3, and T4 clones.
LVMapper seems to be a balanced performer across all types of clones but with a lower precision score.
Regarding precision, SourcererCC, NiCad, and NIL outperform other tools in this category. 
It indicates that Non-LLMs-based methods show strength in detecting T1, T2, and VST3 clones but struggle with more complex types like MT3 and T4.

For LLMs-based detection tools, we first find that the LLaMA-7B and LLaMA2-7B models, which did not undergo instruct-tuning, demonstrate an inability to follow instructions effectively and output meaningful content. 
In contrast, Alpaca, Vicuna, LLaMA2-Chat-7B, and Falcon-Instruct-7B all went through instruction tuning, thus showing high recall results for all types of cloned pairs, albeit with low precision. This suggests that these models may detect all clone pairs as positive, indicating potential shortcomings in accurately detecting cloned code.
LLaMA's report reveals that its code data during training constitutes 4.5\% of the entire training corpus, the lowest amongst all the open-source base models in the experiments. Alpaca and Vicuna-7B, which are fine-tuned based on the LLaMA, did not fine-tune their models on the code task, possibly leading to inferior clone detection capabilities. Besides, LLaMA2's report shows that its code data proportion during training achieves 8.38\%.
Notably, LLaMA2-Chat-7B and Falcon-Instruct-7B have substantially improved recall, but the precision remains relatively low, indicating a high number of false positives. However, their high recall scores suggest they are unlikely to miss any actual clones, making them valuable tools in clone detection tasks where missing a potential clone could have significant consequences.


MPT-Instruct-7B shows a notable ability to detect cloned code with relatively high accuracy. 
MPT \cite{MosaicML2023Introducing} states that its training corpus includes 10\% tokens of code that are cleaned and processed adequately, indicating its competence in handling clone detection tasks.
MPT-INstruct-7B, which is instruct-tuned based on the MPT, implies the capability to follow human instructions to detect code clones.
StarChat-$\beta$'s recall is relatively high compared to the above models with guaranteed precision. Its base model, StarCoderPlus, is trained on over 80 code languages, and it's fine-tuned on the "uncensored" variant of the openassistant-guanaco dataset, specifically constructed for the code task, which could explain its better performance in the code clone detection.
GPT-3.5-Turbo and GPT-4 show the best results in the experiment. This is because these two models have much larger parameters than the other models (GPT-3.5-Turbo contains 175B parameters; falcon contains 7b parameters; StarChat-$\beta$ contains 16b parameters), so they can accommodate more knowledge~\cite{kaplan2020scaling}. The GPT models also have somewhat richer word lists~\cite{openai2023gpt4}, allowing for fine-grained, adequate handling of code tasks.
Moreover, GPT models can handle longer samples without forgetting the opening instruction. In contrast, the other open-source models forget the instruction in many cases and answer incorrectly or even do not follow the instruction in most cases. 

Interestingly, we find that GPT-3.5-Turbo and GPT-4 do not have the same understanding of cloning. GPT-3.5-Turbo focuses on the semantics of the code. Otherwise, GPT-4 combines code structures and semantics into account. For example, a Type-2 clone pair that encodes and decodes a file differs only in the name of the function (\ie encodeFiletoFile, decodeFiletoFile) and the encoding way (\ie Base64.ENCODE, Base64.DECODE). GPT-3.5-Turbo considers the code not to be a clone because the implementation of the functions are mutually exclusive, while GPT-4 considers the code to be a clone because the structure and function of the code fragments are very similar, except that one performs encoding and the other decoding. Due to the imprecise understanding of cloning, the Type-2 recall of GPT-3.5-Turbo is low since the Type-2 code pair is structurally similar, leading the model to output the wrong result.

\begin{titleEnv}
\emph{
\textbf{Summary:}
Using open-source LLMs for clone detection yields superior results in identifying Type-3 and Type-4 clone pairs when relying solely on a simple prompt. However, it does exhibit slightly poorer performance when detecting Type-1 and Type-2 clone pairs compared to existing tools. Notably, GPT-3.5-Turbo and GPT-4 stand out with the highest recall and accuracy rates across nearly all clone types.
}
\end{titleEnv}

\begin{table*}[]
\centering
\caption{Response for RQ2-4 by LLaMA2-Chat-7B.}
\label{tab:rq2-4-example}
\begin{spacing}{1}
\small
\setlength{\tabcolsep}{0.65mm}{

\begin{tabular}{c|c|m{15.2cm}}
\toprule
\toprule
   \textbf{Section} & \textbf{Type}  & \multicolumn{1}{c}{\textbf{Content}} \\
\midrule
\textbf{-} & Nonclone &  \colorbox{codegray}{\textbf{Code1:}} public SocketChannel getChannel() \{\textbackslash n return channel;\textbackslash n\}  \newline \colorbox{codegray}{\textbf{Code2:}} public static void copyFile2(File srcFile, File destFile) throws IOException \{\textbackslash n FileUtils.copyFile(srcFile, destFile);\textbackslash n\} \\
\midrule
\textbf{4.2.4} & Response & Sure, I can help you with that! Here's the analysis of the two code snippets:
\textbf{More similar lines of code:
return channel; and FileUtils.copyFile(srcFile, destFile);}
Both of these lines return a value without any additional logic. They are similar in terms of their simplicity and lack of complexity.
Based on the above analysis, I would say that these two code snippets are a clone pair. \textbf{Yes, they are similar enough to be considered clones}. \\
\bottomrule
\bottomrule
\end{tabular} }%
\end{spacing}
\end{table*}

\subsection{RQ2: Performance of One-Step Chain-of-Thought Prompts \label{rq2exp}}

In this section, we design the prompts using the one-step chain-of-thought to request LLMs to conduct code clone detection from five perspectives. Noted that open-source LLMs do not follow the prompts well. As shown in Table~\ref{tab:rq2-4-example}, we can observe that we request the latest open-source LLM, LLaMA2-Chat-7B~\cite{touvron2023llama2}, to provide the similar lines in the code pair and conduct the code clone detection. However, given the code pair that is completely different in structure as well as semantics, LLaMA2-Chat-7B identifies the code pair as a clone pair by simply analyzing the complexity of the code. For longer code pairs, open-source models are even more limited by input token restrictions and poor long-text modeling capabilities, and more often than not, the answers contain meaningless analysis and erroneous results. Compared with the open source LLMs, GPT-3.5-Turbo and GPT-4 understand instructions more accurately, can perform the tasks in the prompts better, and can make meaningful responses to the multiple prompts we designed, which can more realistically reflect the impact of analyzing the large model from different perspectives on code clone detection. Therefore, we only evaluate GPT-3.5-Turbo and GPT-4 in the following experiments. 

\begin{table}[]
\small
\caption{Recall and Precision on Clone Type Reasoning}
\label{tab:rq2-1}
\begin{spacing}{1.}
\setlength{\tabcolsep}{1mm}{
\begin{tabular}{c|ccccccc}
\toprule
\toprule
\multirow{2}{*}{\textbf{Methods}} & \multicolumn{6}{c}{\textbf{Recall}}   & \multirow{2}{*}{\textbf{Precision}} \\ \cmidrule{2-7}
                      & \textbf{T1}  & \textbf{T2}   & \textbf{VST3}  & \textbf{ST3}  & \textbf{MT3} & \textbf{T4} &                                   \\  
\midrule
GPT-3.5-Turbo               & 1 & 0.98 & 0.98 & 0.94 & 0.87 & 0.36    & 0.77                              \\
\midrule
GPT-4                 & 0.99 & 1 & 0.98 & 0.98 & 0.92 & 0.25   & 0.89                              \\
\bottomrule
\bottomrule

\end{tabular} }%
\end{spacing}
\end{table}

\subsubsection{Clone Type}
\ 
\newline
We request models to analyze the clone type of two code snippets and output the code clone judgment. It is worth noting that we do not inform the models through prompt what are code clone types. The models, including GPT-3.5-Turbo and GPT-4, have prior knowledge of this and can correctly know the four clone types in the clone detection task. From Table~\ref{tab:rq2-1}, we can observe that the recall of MT3 and Type-4 achieves 0.87 and 0.36, respectively. Compared with RQ1, the improvement in Type-2 is huge (i.e., from 0.57 to 0.98) because the clone types mentioned in the prompt help models determine from a more comprehensive perspective. GPT-3.5-Turbo conducts clone detection mainly by analyzing the semantics and neglects to analyze the code structure. When GPT-3.5-Turbo is required to analyze the clone type first in the prompt, the model will consider more structural clones (Type-2 is structural clones). Therefore the clone detection performance of GPT-3.5-Turbo will be greatly improved. For GPT-4, the recall of MT3 and Type-4 achieves 0.92 and 0.25, respectively. These results suggest that having the models analyze the clone type first improves its clone detection overall.

\begin{figure}[htbp]
\centerline{\includegraphics[width=0.35\textwidth]{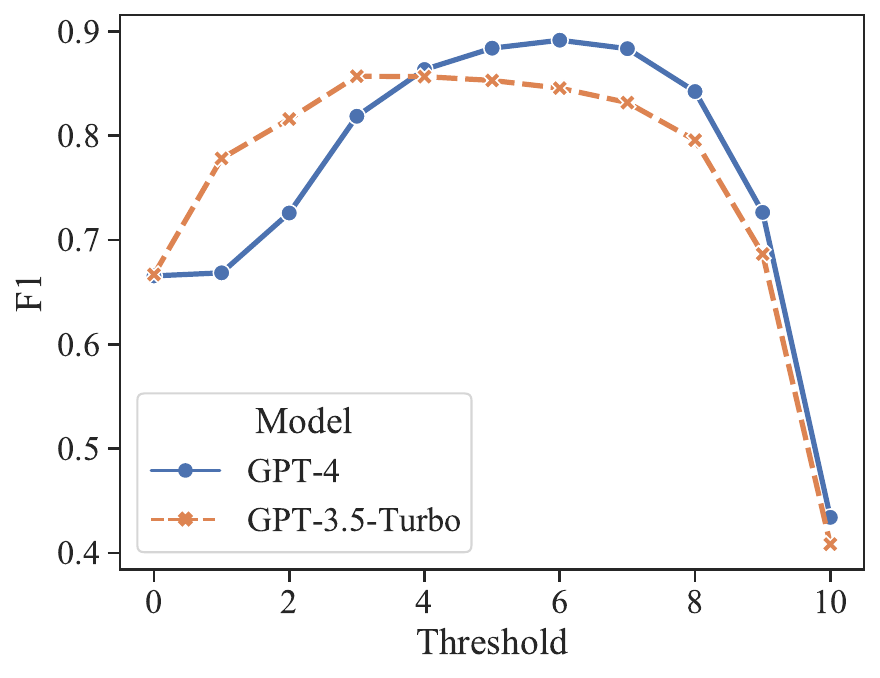}}
\caption{The performance of the two models at different similarity thresholds.}
\label{fig:rq2-2-gpt}
\vspace{-1em}
\end{figure}

\subsubsection{Similarity}
\ 
\newline
We ask the models to output the similarity of the two code snippets instead of outputting the judgment.
By simulating human scoring, we want to assess how well the model understands the cloned code. 
We evaluate the model for code clone detection by setting different thresholds for similarity. 
As shown in Figure \ref{fig:rq2-2-gpt}, we find that the highest F1 value for the GPT-4 is obtained when the similarity threshold is set to six (\ie the precision is 0.93 and the recall is 0.86). And when the similarity threshold for GPT-3.5-Turbo is set to three, the model has the highest F1 value (\ie the precision is 0.86, and the recall is 0.85). Meanwhile, the highest value of F1 for GPT-4 is better than GPT-3.5-Turbo.

\begin{table}[]
\small
\caption{Recall and Precision on Detailed Reasoning}
\label{tab:rq2-3}
\begin{spacing}{1.}
\setlength{\tabcolsep}{1mm}{
\begin{tabular}{c|ccccccc}
\toprule
\toprule
\multirow{2}{*}{\textbf{Methods}} & \multicolumn{6}{c}{\textbf{Recall}}   & \multirow{2}{*}{\textbf{Precision}} \\ \cmidrule{2-7}
                      & \textbf{T1}  & \textbf{T2}   & \textbf{VST3}  & \textbf{ST3}  & \textbf{MT3} & \textbf{T4} &                                   \\  
\midrule
GPT-3.5-Turbo                  & 1 & 0.91 & 0.93 & 0.81 & 0.61 & 0.1   & 0.93                       \\ 
\midrule
GPT-4                    & 0.99 & 1 & 1 & 0.99 & 0.91 & 0.26   & 0.91                         \\ 
\bottomrule
\bottomrule
\end{tabular} }%
\end{spacing}
\vspace{-1em}
\end{table}

\subsubsection{Reasoning}
\ 
\newline
We request models to output the reasoning of the clone detection and, based on the reasoning, output the final judgment. The reasoning process contains how the model comprehends the code as well as the clone detection task. The generated information will be provided to models as additional information to assist models in making judgments and improve the accuracy of clone detection. As shown in Table~\ref{tab:rq2-3}, we find that for GPT-3.5-Turbo, the recall of Type-2 achieves 0.91. As for GPT-4, the recall of MT3 and Type-4 achieve 0.91 and 0.26, respectively.

\begin{table}[]
\small
\caption{Recall and Precision on Similar Line Reasoning}
\label{tab:rq2-4}
\begin{spacing}{1.}
\setlength{\tabcolsep}{1mm}{
\begin{tabular}{c|ccccccc}
\toprule
\toprule
\multirow{2}{*}{\textbf{Methods}} & \multicolumn{6}{c}{\textbf{Recall}}   & \multirow{2}{*}{\textbf{Precision}} \\ \cmidrule{2-7}
                      & \textbf{T1}  & \textbf{T2}   & \textbf{VST3}  & \textbf{ST3}  & \textbf{MT3} & \textbf{T4} &                                   \\  
\midrule
GPT-3.5-Turbo               & 1    & 0.99 & 0.98 & 0.92 & 0.86 & 0.23    & 0.86                              \\ 
\midrule
GPT-4                 & 1 & 1  & 0.99 & 0.99 & 0.88 & 0.26   & 0.90                              \\ 
\bottomrule
\bottomrule
\end{tabular} }%
\end{spacing}
\end{table}

\subsubsection{Similar Line}
\ 
\newline
We request models to output similar lines in the code snippets and, given the similar lines, output the final determination. With the requirement, models are first analyzed in terms of finding similar lines. This perspective differs from direct clone detection in that it requires that models do not need to analyze from the full semantics but can instead make analysis against local code fragments. When the model outputs similar lines, the reasons given can be used as additional information to improve the accuracy of the code for clone detection. As shown in Table~\ref{tab:rq2-4}, we find that for GPT-3.5-Turbo, the recall of Type-2 and MT3 achieves 0.99 and 0.86, respectively, which is a great boost compared with RQ1. As for GPT-4, the recall of MT3 and Type-4 achieves 0.88 and 0.26, respectively.

\subsubsection{Integrated}
\ 
\newline
In this part, we would like to understand the performance of the large model when given multiple-perspective information for clone detection. 
Combining the perspectives from the previous prompts may provide the model with more information, or it may interfere with models since different information may characterize different cloning results, especially when too much information is present. 
As shown in Table~\ref{tab:rq2-5}, we find that for GPT-3.5-Turbo, the recall of Type-2 achieves 0.95. However, compared with the former prompts, the other recall results have a huge decrease, and the recall of MT3 and Type-4 is even lower than the original result in RQ1. 
For GPT-4, the recall of MT3 and Type-4 achieve by 0.91 and 0.32, respectively. This indicates that GPT4 outperforms GPT-3.5-Turbo in terms of understanding and analyzing inputs with complex information perspectives and long texts.


\begin{titleEnv}
\emph{
\textbf{Summary:}
The clone detection performance of GPT-3.5-Turbo and GPT-4 can be improved by requiring models to provide clone type, similarity, reasoning, and similarity lines.
Using one-step chain-of-thought prompts allows the models to analyze code pairs and intermediate reasoning, leading to better clone detection.
}
\end{titleEnv}

\begin{table}[]
\small
\caption{Recall and Precision on Integrated Reasoning}
\label{tab:rq2-5}
\begin{spacing}{1.}
\setlength{\tabcolsep}{1mm}{
\begin{tabular}{c|ccccccc}
\toprule
\toprule
\multirow{2}{*}{\textbf{Methods}} & \multicolumn{6}{c}{\textbf{Recall}}   & \multirow{2}{*}{\textbf{Precision}} \\ \cmidrule{2-7}
                      & \textbf{T1}  & \textbf{T2}   & \textbf{VST3}  & \textbf{ST3}  & \textbf{MT3} & \textbf{T4} &                                   \\  
\midrule
GPT-3.5-Turbo                  & 0.89 & 0.95 & 0.88 & 0.8 & 0.58 & 0.07    & 0.97                       \\ 
\midrule
GPT-4                    & 1 & 1 & 0.99 & 0.98 & 0.91 & 0.32   & 0.90                 \\ 

\bottomrule
\bottomrule
\end{tabular} }%
\end{spacing}
\end{table}

\subsection{RQ3: Performance of Multi-Step Chain-of-Thought Prompts}
\subsubsection{Separate Explanations}
\ 
\newline
In this section, we aim to assess the impact of four types of independent intermediate reasoning (RQ2) on clone detection. We independently ask models to explain the code from each of the four perspectives in RQ2 and yield the corresponding intermediate reasoning. The prompts in this section differ from those in RQ2 in that the latter generate intermediate reasoning, which may be based on other reasoning. However, in this section, the generating process is independent of each other, and every time models are asked a question, they are asked in a new context. Subsequently, we combine the four types of intermediate reasoning into a prompt and task the models with performing the clone detection. Table~\ref{tab:rq3-1} presents that for GPT-3.5-Turbo, the recall of MT3 and Type-4 achieves 0.92 and 0.39, respectively. Compared with the RQ2-5, the recall of MT3 and Type-4 increased by 0.34 and 0.32, respectively. These findings suggest that GPT-3.5-Turbo cannot effectively analyze multiple interacted intermediate reasoning, which hinders its ability to determine clone detection accurately. For GPT-4, compared with RQ2-5, the recall, and precision do not vary much, indicating that GPT-4 demonstrates superior capability in comprehending and utilizing the four intermediate reasoning to boost clone detection.

\subsubsection{Separate Codes}
\ 
\newline
In this section, we aim to replicate the current deep-learning-based clone detection techniques that characterize code features independently to predict outcomes. Therefore, we first request GPT-3.5-Turbo and GPT4 to generate independent code explanations to characterize the code and then input these explanations to models for clone detection.
To ensure independent and unbiased explanations, we first divide the code pairs and ask the models to explain each code separately. This prevents any influence during the generation of the code explanations. Then, we combine the codes and their explanations into a prompt and ask the models to perform code clone detection. By separately analyzing the two pieces of code, we prevent any interference and aim to evaluate the performance of clone detection when the model is also given independent code explanations. From Table~\ref{tab:rq3-2}, we find that for GPT-3.5-Turbo, the precision achieves 0.9, and the recall of MT3 and Type-4 achieves 0.76 and 0.19, respectively. For GPT-4, the precision achieves 0.96, and the recall of MT3 and Type-4 achieves 0.83 and 0.29. It indicates that compared with RQ1, multi-step chain-of-thought reasoning by separating codes can improve the performance of the clone detection for GPT-3.5-Turbo and GPT-4.

\begin{titleEnv}
\emph{
\textbf{Summary:}
The clone detection performance of GPT-3.5-Turbo and GPT-4 can be improved by Multi-Step Chain-of-Thought prompts, including separating explanations and codes. Different from RQ2, separating explanations provide models of independent intermediate reasoning of code, and separating codes provide models of independent explanation of code, which avoid the influences between generated information.
}
\end{titleEnv}

\begin{table}[]
\small
\caption{Recall and Precision on Separate Explanations}
\label{tab:rq3-1}
\begin{spacing}{1.}
\setlength{\tabcolsep}{1mm}{
\begin{tabular}{c|ccccccc}
\toprule
\toprule
\multirow{2}{*}{\textbf{Methods}} & \multicolumn{6}{c}{\textbf{Recall}}   & \multirow{2}{*}{\textbf{Precision}} \\ \cmidrule{2-7}
                      & \textbf{T1}  & \textbf{T2}   & \textbf{VST3}  & \textbf{ST3}  & \textbf{MT3} & \textbf{T4} &                                   \\  
\midrule
GPT-3.5-Turbo                  & 1 & 0.98 & 0.97 & 0.95 & 0.92  & 0.39   & 0.79                       \\ 
\midrule
GPT-4                    & 1   & 0.99   & 1    & 0.99    & 0.93    & 0.33      & 0.90                          \\ 
\bottomrule
\bottomrule
\end{tabular} }%
\end{spacing}
\vspace{-1em}
\end{table}

\begin{table}[]
\small
\caption{Recall and Precision on Separate Codes}
\label{tab:rq3-2}
\begin{spacing}{1.}
\setlength{\tabcolsep}{1mm}{
\begin{tabular}{c|ccccccc}
\toprule
\toprule
\multirow{2}{*}{\textbf{Methods}} & \multicolumn{6}{c}{\textbf{Recall}}   & \multirow{2}{*}{\textbf{Precision}} \\ \cmidrule{2-7}
                      & \textbf{T1}  & \textbf{T2}   & \textbf{VST3}  & \textbf{ST3}  & \textbf{MT3} & \textbf{T4} &                                   \\  
\midrule
GPT-3.5-Turbo                  & 0.98 & 0.97 & 0.92 & 0.87 & 0.76 & 0.19   & 0.90                       \\ 
\midrule
GPT-4                    & 1  & 0.98 & 0.95 & 0.97 & 0.83 & 0.29   & 0.96                       \\ 

\bottomrule
\bottomrule
\end{tabular} }%
\end{spacing}
\end{table}

\subsection{RQ4: Performance of Code Embedding}

This section offers a comparative analysis of the performance of various LLMs, specifically focusing on their usage of code embedding. 
This is done by contrasting their results with established PLMs, including CodeBERT-base, CodeBERT-mlm, and text-embedding-ada-002.
In our study, the performance of three models, namely CodeBERT, CodeBERT-MLM, and Text-embedding-ada-002, was evaluated based on their ability to identify cloned code pairs in the BigCloneBench dataset.
The models were trained to predict similarity between pairs of code, which was computed as the cosine similarity between the vector representations of the code pairs.

\begin{figure}
\centering
\begin{subfigure}{.238\textwidth}
  \centering
  \includegraphics[width=\linewidth]{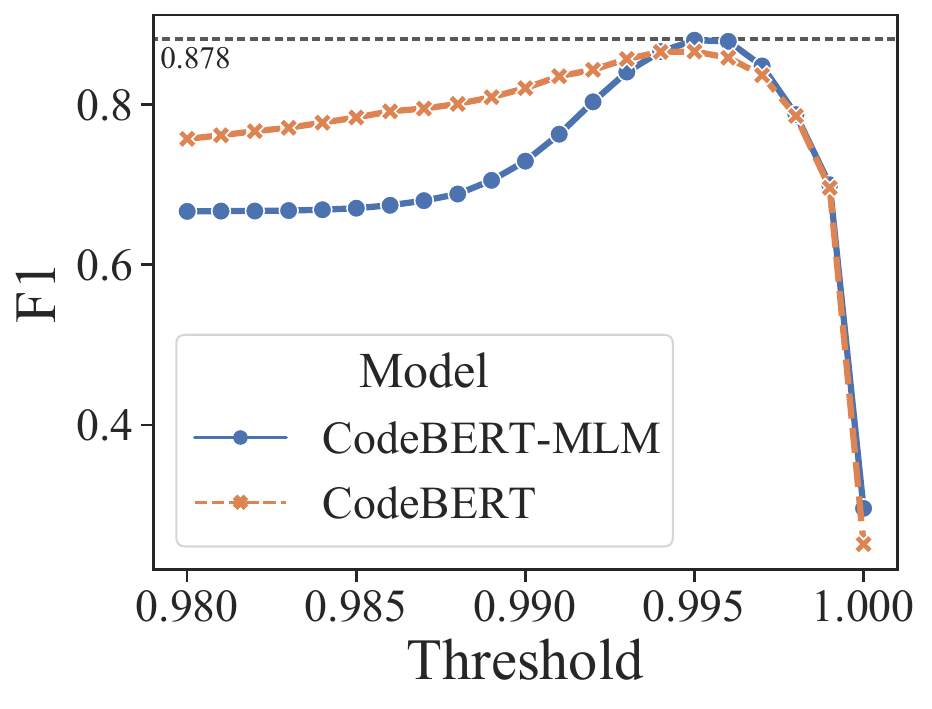}
  \label{fig:sub1}
\end{subfigure}%
\begin{subfigure}{.238\textwidth}
  \centering
  \includegraphics[width=\linewidth]{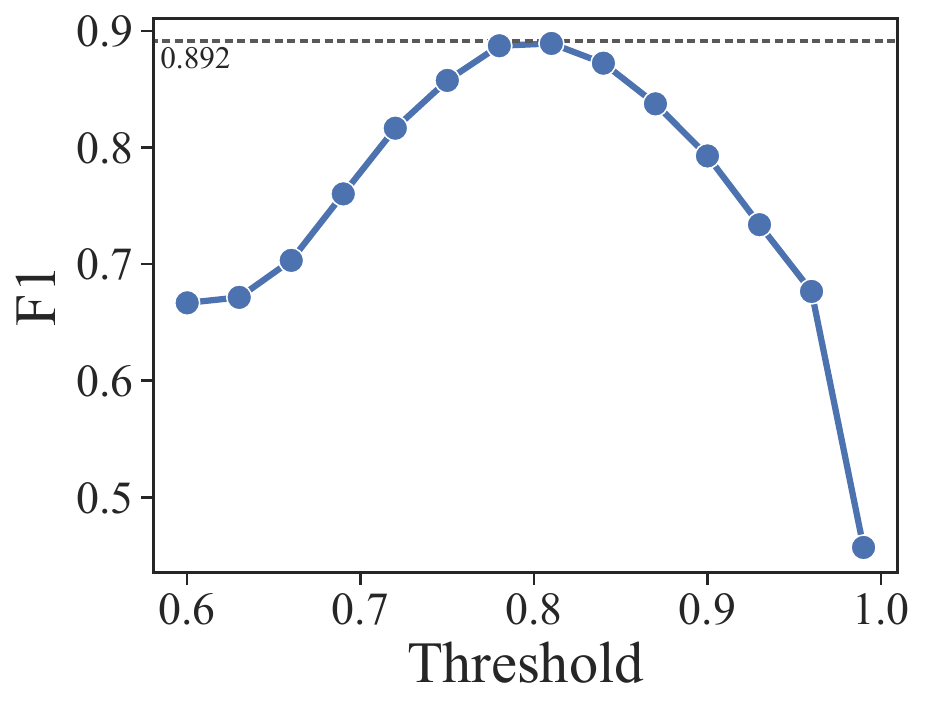}
  \label{fig:sub2}
\end{subfigure}
\vspace{-2.5em}
\caption{The left figure shows the F1 performance of CodeBERT and CodeBERT-MLM at different thresholds. The right figure shows the performance of Text-embedding-ada-002.}
\label{fig:rq4-embedding}
\vspace{-1em}
\end{figure}

\begin{figure}
\centering
\begin{subfigure}{.238\textwidth}
  \centering
  \includegraphics[width=\linewidth]{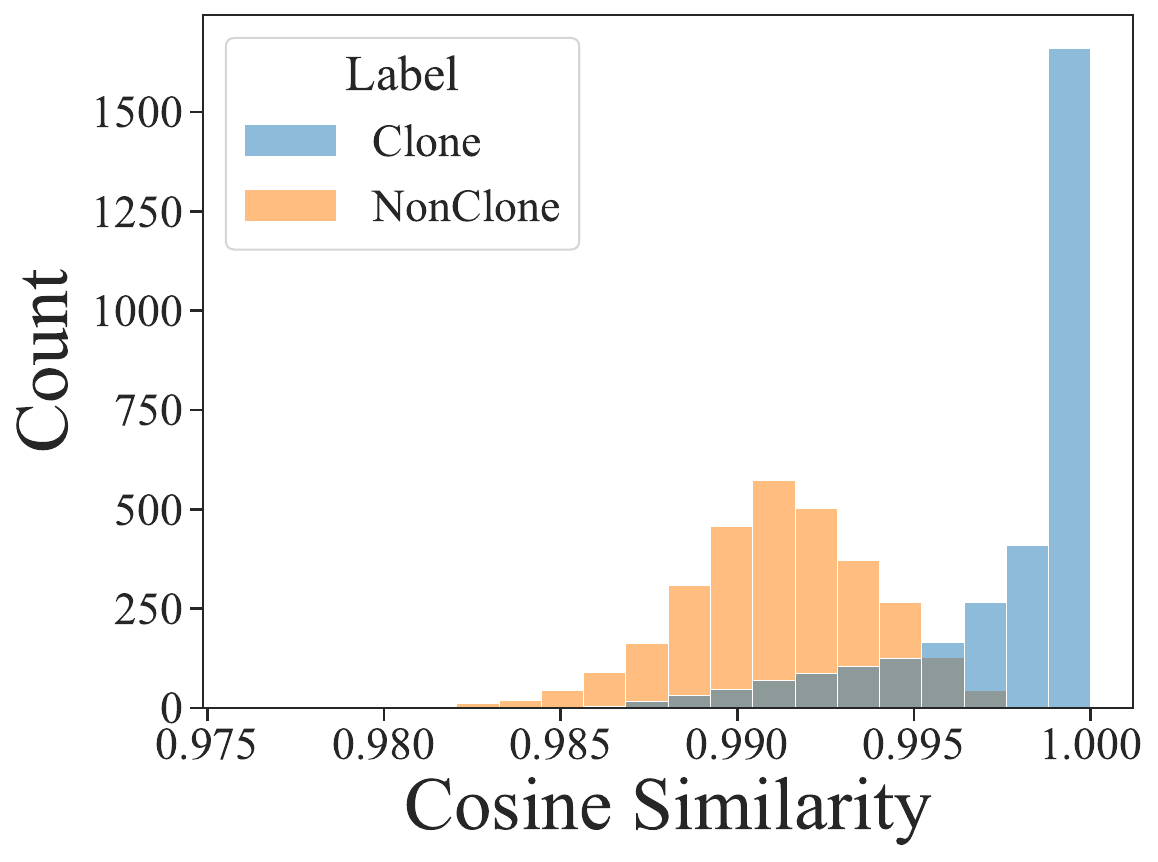}
  \label{fig:sub1}
\end{subfigure}%
\begin{subfigure}{.238\textwidth}
  \centering
  \includegraphics[width=\linewidth]{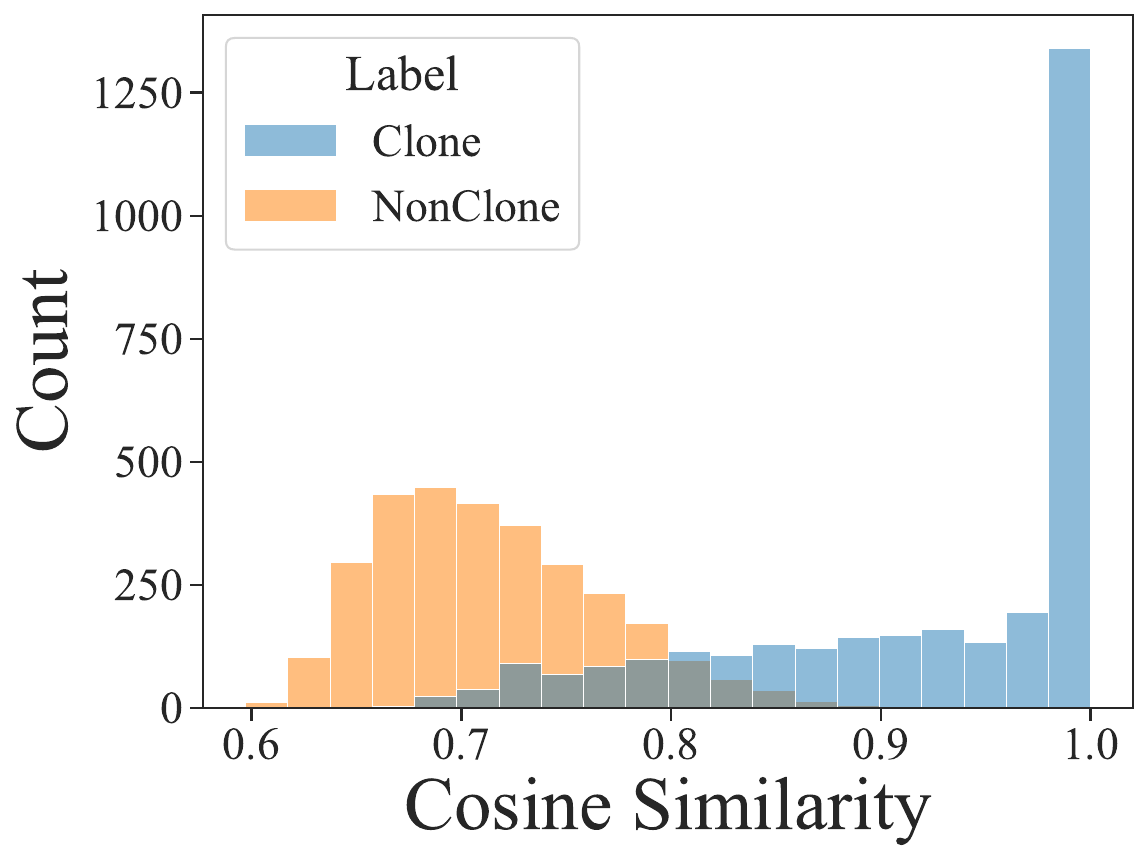}
  \label{fig:sub2}
\end{subfigure}
\vspace{-2.5em}
\caption{The left figure shows the similarity distribution between the two codes embedded by CodeBERT-MLM. The right figure shows the distribution of similarity between the two codes embedded by Text-embedding-ada-002.}
\label{fig:rq4-distribution}
\end{figure}

In order to capture the nuances of model's performance, we varied the probability thresholds and measured the precision, recall, and F1 scores at each level.
Each model was analyzed at its respective threshold which corresponded to its optimal performance. The comparative F1-score across the different thresholds is graphically represented in Figure \ref{fig:rq4-embedding}.
The models' optimal performance was observed at different threshold levels: 0.995 for both CodeBERT-base and CodeBERT-MLM, and 0.8 for Text-embedding-ada-002. 
These thresholds were identified based on the peak performance of each model under evaluation. 
While all models demonstrated strong performance in several categories, they exhibited reduced effectiveness in WT3/T4 and NoClone scenarios.
Interestingly, CodeBERT-MLM surpassed CodeBERT-base at its peak performance, showing superior outcomes in MT3, ST3, and VST3 scenarios. 
However, Text-embedding-ada-002 outperformed both CodeBERT variants, showcasing the highest precision and F1 score, thereby demonstrating robust performance even at a lower threshold.

Specifically, Text-embedding-ada-002 achieved the highest overall F1 score. 
As illustrated in Figure~\ref{fig:rq4-distribution}, this model exhibited a more expansive range of similarity scores, enabling a more effective distinction between true and false positives. 
This broader distribution, however, also resulted in a few mispredictions at higher similarity scores. 
Despite these occasional high-similarity mispredictions, the findings strongly suggest that Text-embedding-ada-002 provides the most robust performance in the detection of cloned code pairs. 
Its larger distribution range of similarity scores further substantiates its reliability and effectiveness in differentiating between cloned and non-cloned code pairs, thereby showcasing the model's robustness.


\begin{titleEnv}
\emph{
\textbf{Summary:}
Text-embedding-ada-002 is more effective than specialized CodeBERT models in identifying cloned code, exhibiting superior overall performance. 
The advantage of Text-embedding-ada-002 lies in its capacity to generate a wider range of similarity scores, leading to better discrimination between true and false positives. 
}
\end{titleEnv}

\subsection{RQ5: Performance Across Different Programming Languages}

In this section, we analyze the performance of LLMs in detecting code clones across different programming languages. 
Evaluating these models unveils a pattern wherein they display remarkable precision, as evidenced by the data presented in Tables \ref{tab:rq5}. 
The superior recall rate of GPT-4 across all clone types and languages, especially Python and C++, suggests that GPT-4's improved code clone detection capacity may be ascribed to an advanced understanding of various syntax and structures across these programming languages. 

\begin{table}[htbp]
\small
  \centering
  \caption{Recall and Precision on Java, Python, and C++ Code Clone Detection}
    \setlength{\tabcolsep}{0.88mm}
    \begin{tabular}{cc|ccccccc}
    \toprule
    \toprule
    \multicolumn{2}{c|}{\multirow{2}[4]{*}{\textbf{Methods}}} & \multicolumn{6}{c}{\textbf{Recall}}           & \multirow{2}[4]{*}{\textbf{Precision}} \\
\cmidrule{3-8}    \multicolumn{2}{c|}{} & \textbf{T1} & \textbf{T2} & \textbf{VST3} & \textbf{ST3} & \textbf{MT3} & \textbf{T4} &  \\
    \midrule
    \multirow{3}[2]{*}{GPT-3.5-Turbo} & Java  & 1     & 0.57  & 0.85  & 0.78  & 0.59  & 0.09  & 0.95 \\
          & Python & 0.99  & 0.94  & 0.61  & 0.46  & 0.41  & 0.22  & 0.99 \\
          & C++   & 0.99  & 0.99  & 0.68  & 0.44  & 0.33  & 0.16  & 1 \\
    \midrule
    \multirow{3}[2]{*}{GPT-4} & Java  & 1     & 0.98  & 0.99  & 0.94  & 0.77  & 0.15  & 0.96 \\
          & Python & 1     & 0.99  & 0.99  & 0.99  & 0.9   & 0.72  & 1 \\
          & C++   & 1     & 1     & 0.97  & 0.95  & 0.87  & 0.67  & 1 \\
    \bottomrule
    \bottomrule
    \end{tabular}%
  \label{tab:rq5}%
  \vspace{-1em}
\end{table}%

These differences across Python, C++, and Java might also be attributed to the inherent complexity of each language's syntax and structure. Python's simplicity and high-level abstraction might make clone detection relatively more straightforward, reflected in the impressive performance of both models. The high recall values for Python could also be influenced by the volume of Python code available during the training phase of the models, as Python is one of the most commonly used languages in software development and AI research. Also, it is plausible that the datasets used for evaluating Python and C++ clone detection might overlap with the training data of the LLMs, leading to a seemingly better performance.

\begin{titleEnv}
\emph{
\textbf{Summary:}
The performance of LLMs in code clone detection varies across different programming languages, with a trend of superior results in Python, likely due to its inherent simplicity and prevalence in training data. 
}
\end{titleEnv}

\section{Discussions and Limitations}


\subsection{Discussions}

\subsubsection{\textbf{Does the use of CoT improve LLMs' clone code detection capabilities universally?}}
The use of CoT has been known to enhance performance on complex tasks like mathematical reasoning. 
However, our study suggests that the implementation of CoT does not necessarily lead to an overall improvement in LLMs' clone code detection capabilities.
There are two prime reasons for this case.
First, the necessity for LLMs to have a strong ability to follow human instructions. Models without instruction tuning may fail to improve performance through CoT as they lack the capacity to follow human instructions effectively.
Second, given the complexity of clone code detection tasks, particularly due to the requirement of matching two sets of code, LLMs need robust abilities in long document reasoning and context understanding. 
If these capabilities are lacking in the LLMs, they may fail to comprehend human instructions and even underperform without the implementation of CoT. 
Therefore, while CoT can enhance the clone code detection abilities of more capable LLMs, it may detrimentally affect the performance of weaker models. 

\subsubsection{\textbf{Why does CoT enhance the performance of stronger LLMs in clone detection?}}
CoT enhances the performance of stronger LLMs in clone detection by extending the context of the model's prediction process. 
In a normal scenario without CoT, the model responds based solely on the two given code samples. 
However, when CoT is implemented, the context for predicting the response tokens includes not only the two code samples but also the model's own thought process.
This offers a more comprehensive analysis of the code pair and subsequently enhances the model's clone code detection performance.

\subsubsection{\textbf{Why does code embedding perform better than LLMs chat in clone detection tasks?}}
The success of code embedding over LLMs chat is attributed to its different approach to detecting cloned code. 
Code embedding creates an individual representation for each code through a text encoder, which is then compared using cosine similarity. 
This process does not involve a comparative analysis of the two codes, thereby simplifying the task as compared to directly performing clone detection on two codes. 
Although LLMs with CoT can provide an analysis for each code and then compare the results, the output in the form of natural language text makes this process more complex compared to direct encoding to obtain representations. 
Also, the final comparison stage still demands a strong context-understanding capability from LLMs to compare the longer code segments.
As a result, the code embedding task appears simpler both in terms of code analysis and code pair comparison, thereby leading to better performance.

\subsubsection{\textbf{Open source and expenses.}}
To contribute to the academic community and promote further advancements in code clone detection research, we will make publicly available the inference results of GPT-3.5-Turbo and GPT-4 on Java, C++, and Python. 
In addition, we will release a meticulously curated dataset consisting of over 200,000 clone pairs for Python and C++, each classified into clone types: VST3 clones, ST3 clones, MT3 clones, and WT3/T4) clones. 
We believe these resources will substantially facilitate future exploration and development in this area.
All the data are available at the link\footnote{\href{https://github.com/LLM4CodeClone/LLM4CodeClone}{https://github.com/LLM4CodeClone/LLM4CodeClone}}.
Furthermore, in this study, we spent over \textbf{\$3500} on the OpenAI APIs queries of GPT-3.5-Turbo and GPT-4.
The entirety of the experiments included prompt design consumption, chain-of-thought experiments, and multi-language experiments, using a total of \textbf{6,942,335} tokens.

\subsection{Limitations}


\subsubsection{\textbf{Limitations in constructing the instruction set.}}
Although we have constructed a set of instructions based on a small sample size, these may not necessarily be optimal for clone code detection tasks. 
Determining the most effective instructions would require extensive trials, which can be prohibitive due to resource constraints.
We aim to address this limitation in future work by exploring a broader set of instructions for this task.

\subsubsection{\textbf{Selection of models for evaluation.}}
The field of LLMs is ever-evolving with the frequent introduction of new models. 
In the current study, our selection of models was limited to a subset of these, chosen based on factors such as their novelty, popular usage, and established performance in neural language tasks. 
Additionally, due to computational resource limitations, we were restricted to testing models at the 7B and 16B parameter scales, which, though impressive in the context of software engineering, still leaves room for exploration. 
In future studies, we intend to extend our evaluation to include a broader range of models and larger scales to provide a more comprehensive understanding of the capabilities of LLMs in software engineering.

\subsubsection{\textbf{Absence of demonstrations during in-context learning.}}
Our current study does not leverage demonstrations, \ie the use of few-shot examples, known to enhance the performance of LLMs.
This omission was primarily due to resource constraints, as the inclusion of demonstrations can significantly increase the demand for computing resources. 
Additionally, the models tested in this study were on the smaller end of the scale, inherently limiting their capacity for contextual memory. 
Future work will look into incorporating demonstrations and testing larger open-source LLMs, which are expected to have more robust contextual memory capabilities, thereby potentially improving the effectiveness of clone code detection.

\subsubsection{\textbf{Enforcing a response structure during detection.}}
In our detection task, we mandated that the model's response contain either `yes' or `no'. 
However, some models may not adhere to this instruction, leading to potential inconsistencies in evaluation. 
For this assessment, we combined the use of regular expressions with manual checking to determine the correctness of a model's response.
In future studies, we plan to explore more effective evaluation methods or optimize prompts to reduce the reliance on manual checking and accurately assess model responses.

\section{Conclusion}

This study presented a comprehensive empirical evaluation of Large Language Models (LLMs) for automated code clone detection across diverse clone types, languages, and prompt formulations.
The key findings demonstrate that advanced LLMs like GPT-3.5-Turbo and GPT-4 can achieve remarkably high recall and accuracy in detecting even complex semantic clones, outperforming existing techniques. 
Introducing intermediate reasoning steps through chain-of-thought prompting leads to noticeable gains by equipping models with a structured thought process. 
Additionally, representing code as vector embeddings enables effective clone detection, with text encoders like Text-embedding-ada-002 producing superior results over specialized models.
Our study provides strong evidence that LLMs hold significant promise for clone detection by leveraging their natural language proficiency. 
The insights gained will guide future research toward developing more robust LLM-based techniques to enhance software engineering. 
The prompts and evaluation methodologies presented also contribute a useful benchmark for further studies in this emerging domain.

\bibliographystyle{ACM-Reference-Format}
\bibliography{ICSE2024}
\end{document}